\documentclass[amsmath,amssymb,prl,twocolumn]{revtex4}

\usepackage{graphicx}

\begin{document}

\title{Density waves and Cooper pairing on the honeycomb lattice}
\author{Carsten Honerkamp}
\affiliation{Theoretical Physics, Universit\"at W\"urzburg, D-97074 W\"urzburg, Germany 
}
\date{November 8, 2007}

\begin{abstract}
Motivated by the surge in research activities on graphene, we investigate 
instabilities of electrons on the honeycomb lattice, interacting by onsite and nearest-neighbor terms, using a renormalization group scheme. Near half band-filling, critical minimal interaction strengths are required for instabilities toward antiferromagnetic or charge-density wave order. Away from half filling,  $f$-wave triplet pairing and $d+id$ singlet pairing instabilities are found to emerge out of density-wave regimes.
\end{abstract}

\maketitle
 
Recently it has become possible to fabricate graphene\cite{geim}. Already the structure of graphene poses some interesting questions to experiment and theory\cite{geimnovoselov}. Concerning the electronic properties, the massless-Dirac  spectrum near the two Fermi points of the undoped system gives rise to novel quantum Hall effects and many more complex physical puzzles\cite{geimnovoselov}. 
Unconventional mesoscopic properties of graphene as specular Andreev reflection\cite{beenakker}, analogues to relativistic quantum mechanics\cite{kleinparadox} or spin quantum bits with graphene\cite{trauzettel} are likewise discussed intensively.

Many-body effects are expected to play a role in graphene, although no drastic anomalies have been reported yet\cite{rothenberg}. Superconductivity in graphite compounds is usually attributed to extrinsic causes, e.g. to interlayer states for intercalated compounds\cite{C6} or, more intriguing and less clear-cut, to disorder with sulfur atoms in graphite-sulfur composites\cite{kopelevych}.
In theory, the electronic selfenergy for the undoped case was found to be of marginal Fermi liquid type\cite{guinea}. In the doped case, a normal Fermi liquid develops\cite{dassarma}. The Dirac spectrum should further be reflected in the dependence of the phonon renormalization on the electron density\cite{castroneto}. Recently it has been proposed that a strong second-nearest-neighbor repulsion may lead to a nontrivial insulating state with a quantized spin Hall conductance\cite{raghu}. 
There have been various theoretical studies of possible ordered phases on the honeycomb lattice\cite{sorella,khveshchenko,herbut,uchoa,blackschaffer} driven by interactions, but to date no comprehensive picture for a wider range of interaction parameters and band fillings is available. 
Here we investigate the intrinsic interaction effects of electrons on a honeycomb lattice for short-ranged interactions, using a perturbative functional renormalization group (fRG) technique. The fRG is known to give an unbiased picture of the leading ordering tendencies of interacting fermions at low temperatures. Recently it has been applied to two-dimensional (2D) Hubbard models on square\cite{zanchi,halboth,hsfr,tflow} and triangular lattices\cite{tsai,tria}. 

Our model is a 2D honeycomb lattice with nearest-neighbor hopping amplitude $t$. The interaction terms contain onsite and nearest-neighbor repulsions $U$ and $V$, and a spin-spin interaction $J$. The restriction to short-range terms is partly due to the difficulty to treat a long-range Coulomb part directly in the fRG approach. 
However, most experimental graphene systems are doped to some degree. Hence at least the effective interaction is screened, and our starting point may not be unrealistic. Furthermore, for the undoped case, the long-range part was shown\cite{herbut} to be marginally irrelevant in $1/N$, hence our results for short-range interactions may even be useful in the undoped case. The Hamiltonian reads
\begin{eqnarray*} 
H&=& - t \sum_{\langle i,j\rangle , s} \left( c_{i,s}^\dagger c_{j,s} + c_{j,s}^\dagger c_{i,s} \right)  + U \sum_i n_{i,\uparrow} n_{i, \downarrow}  \\  &&\hspace{1.cm} + V \sum_{\langle i,j\rangle, \, s,s'} n_{i,s} n_{j,s'} + J  \sum_{\langle i,j\rangle} \vec{S}_{i}\cdot \vec{S}_{j}  \, . \end{eqnarray*}
$\langle i,j\rangle$ denotes all pairs of neighbored sites and $\vec{S}_i =\frac{1}{2} \sum_{ss'} c_{i,s}^\dagger \vec{\sigma}_{ss'}c_{i,s'}$.

The fRG scheme used here is an approximation to an exact flow equation for the one-particle irreducible vertex functions of a many-fermion system when a parameter in the quadratic part of the action is varied\cite{fRG}. 
In the temperature-flow scheme employed here, the temperature $T$ is used as flow parameter. The fRG flow is generated by lowering $T$ from an initial value $T_0$ where the interaction effects are  negligible.
In the approximation we use\cite{tflow}, the change of the interaction vertex $V_T$ with is given by the $T$-derivative of one-loop particle-hole diagrams, including vertex corrections and screening, and particle-particle diagrams, of second order in the vertices $V_T$. Higher loop contributions are generated by the integration of the flow. Like in many previous works using this method, the self-energy feedback on the flow of $V_T$ is neglected. It may become important when the interactions get large. Hence the flow is stopped when the interaction strength exceeds twice the bandwidth. The method is controlled in the limit of interaction strength going to zero. For the interesting case of moderate interactions, it should be viewed as a step beyond the meanfield level that captures the evolution and competition of various correlations in an unbiased way.

\begin{figure}

\includegraphics[width=.46\textwidth]{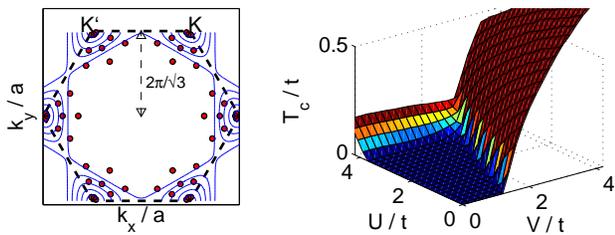} 

\caption{(color online). Left: Brillouin zone and 18$\times 3$ points used for discretizing the wave-vector dependent interaction. The solid lines are at constant band energy. The lattice constant (minimal distance between two A-sublattice sites) is set to unity.  
Right: Critical temperature $T_c$ for the flow to strong coupling vs. interaction parameters $U$ and $V$ at half filling $\mu=0$. In the region with $T_c=0$, the semimetal is stable.
For small $U$ and $V>1.2t$, the flow is toward a CDW instability, for small $V$ and $U>3.8t$ toward a SDW instability. }
\label{Thexplot}
\end{figure}

The interaction vertex can be expressed by a coupling function $V_T (k_1,k_2,k_3)$. It depends on the generalized wavevectors of two incoming particles ($k_1$ and $k_2$) and one outgoing ($k_3$) particle with wavevector, Matsubara frequency and spin projection $k_i=(\vec{k}_i, \omega_i,s_i)$. In the search for instabilities toward symmetry breaking, the frequency dependence is neglected and the $\omega_i$ are set to zero. The $\vec{k}$-dependence of remaining function  $V_T (\vec{k}_1,\vec{k}_2,\vec{k}_3)$ is discretized in the so-called $N$-patch scheme, introduced in this context by Zanchi and Schulz\cite{zanchi}. For standard many-fermion systems with a FS, this amounts to keeping  $V_T (\vec{k}_1,\vec{k}_2,\vec{k}_3)$ constant within patches labeled by $k_i=1,\dots N$ perpendicular to the FS. This defines an $N^3$-component coupling function $V_T (k1_,k_2,k_3)$, which is computed for $\vec{k}(k_i)=\vec{k}_i$, $i=1,\dots 3$, on the FS. As for the honeycomb lattice there is no FS at half band filling, we generalize the patching scheme to two or three rings of 18 or 24 patches around the Dirac points (see Fig.~\ref{Thexplot}). In addition we have a band index for the incoming and outgoing particles.

The RG flow is started at an initial temperature $T_0$. The initial interaction is given by the bare interaction with the onsite repulsion $U$ and the nearest-neighbor repulsion $V$.
Specifically, we search for {\em flows to strong coupling}, where for a certain low temperature $T_c$ one or several components of $V_T (k_1,k_2,k_3)$ become large. At this point the approximations break down, and the flow has to be stopped. Information on low-temperature state is obtained by analyzing which coupling functions grow most strongly and from the flow of susceptibilities. In particular for 2D systems, this instability does not guarantee true long-range order. Rather, it should be interpreted as a breakdown of the (semi-)metallic state and as indicator for the leading correlations at low $T$.    

We begin with the semimetal for zero doping and chemical potential $\mu=0$. The first finding is perturbative stability. Starting the RG flow at high $T$ and small $U$, $V$ or $J$, we can follow the flow down to lowest $T$ without a divergence. 
This is quite different from many-fermion systems with a finite density of states at the Fermi energy, where the flow practically always leads to some kind of instability. In our case, the absence of a flow to strong coupling indicates the absence of long-range order due to electronic interactions even at lowest $T$.

Next we increase the bare interactions. Above a critical value $U_c \sim 3.8t$ for the onsite repulsion $U$ with $V=J=0$, the interactions flow to strong coupling. 
The static antiferromagnetic (AF) spin susceptibility grows most strongly toward the critical temperature scale, indicating a tendency toward AF spin-density wave (SDW) formation with opposite orientation of the ordered spin moment on $A$ and $B$ sublattices. 
For small $U$, $J=0$, and increasing nearest-neighbor repulsion $V$ we again find a flow to strong coupling for $V>V_c \sim 1.2t$, now with leading charge-density wave (CDW) correlations for different charge densities on the two sublattices. 
This compares favorably with a previous $1/N$-analysis\cite{herbut} finding the same instabilities beyond critical values $U_c$ and $V_c$. There is also good agreement with the early Quantum Monte Carlo work by Sorella\cite{sorella} who found a transition to an AF Mott-state at $U \sim 4t$. In Fig.~\ref{Thexplot} we show the dependence of the critical temperature $T_c$ for the flow to strong coupling on the interaction parameters. 
The critical $U$ for boundary of the SDW instability is more or less unaffected by an increasing $V$, while $V_c$ for the CDW regime is shifts to larger $V$ with a roughly linear dependence on $U$. When the two lines meet, there is a continuous change in the flow from leading SDW to leading CDW correlations (or vice versa). This is consistent with a first order transition, if order is possible at all. The competition for the low energy spectral weight not included in this study as the selfenergy is neglected could however reduce the ordered moments in the transition region. 

\begin{figure}
\includegraphics[width=.48\textwidth]{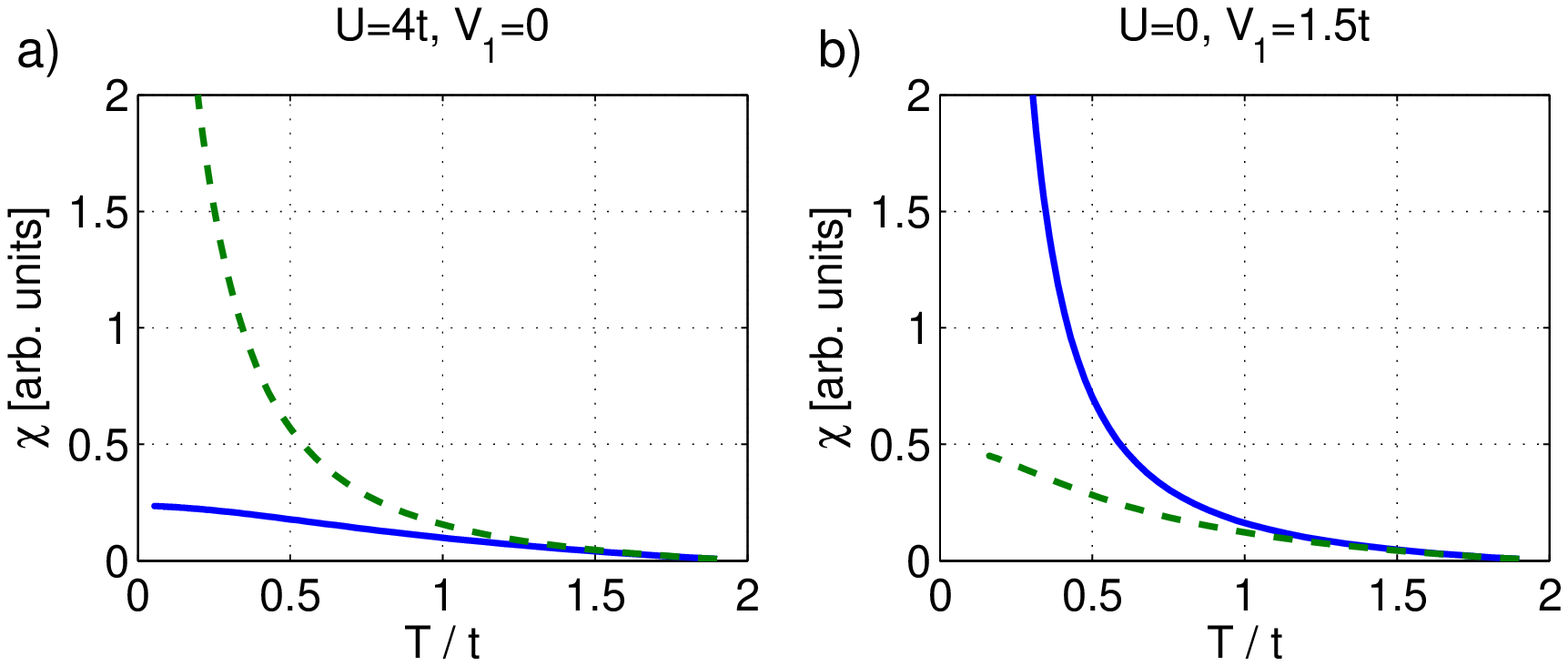}

\includegraphics[width=.48\textwidth]{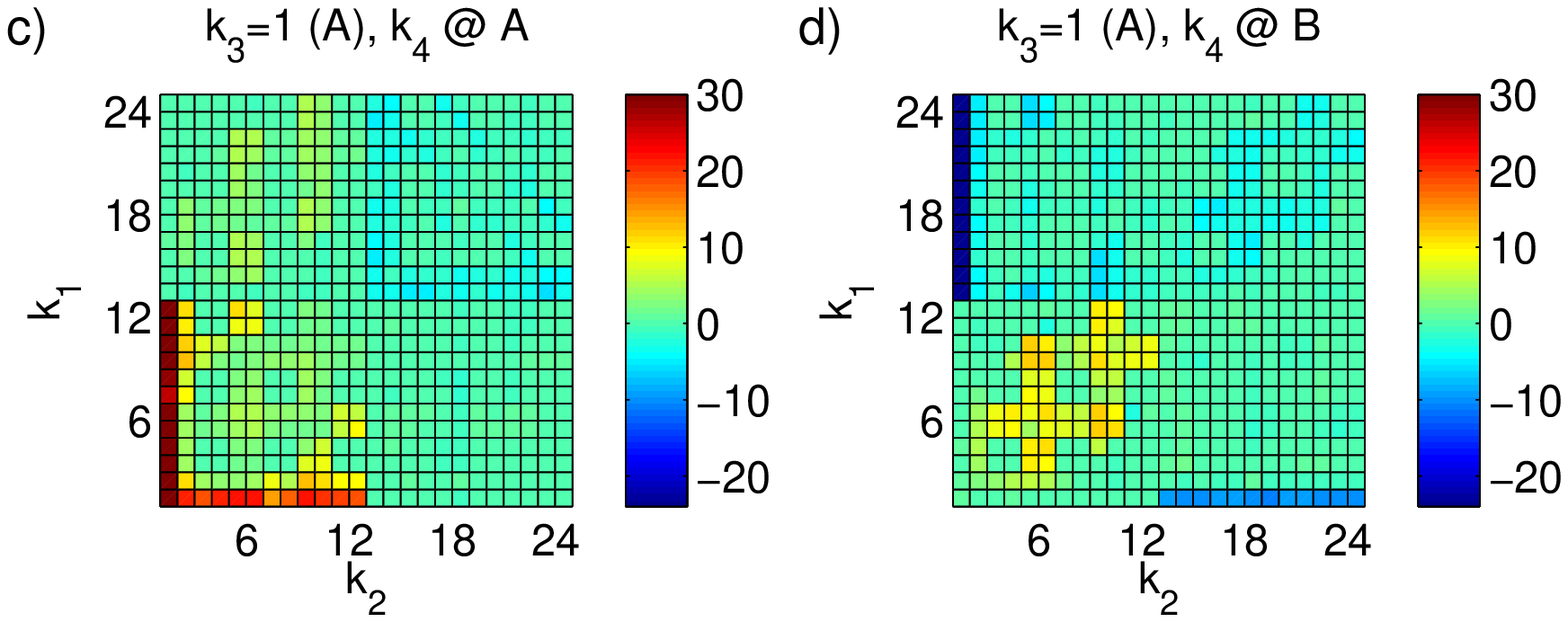}

\caption{(color online). 
Upper plots: Flow of susceptibilities, solid line CDW, dotted line SDW, 
for a) $U$$=$$4t$, $V$$=$$0$, b) $U$$=$$0$, $V$$=$$1.5t$. 
Lower plots: Effective interactions $V_T(k_1,k_2,k_3)$ very close to the SDW instability at $T_c\approx 0.05t$ for $U$$=$$4t$, $V$$=$$0$. The colorbar indicates the values of the couplings. The incoming wavevectors $k_1$ and $k_2$ are on 12 points on the inner rings near the Dirac points, for $k_{1/2}=$1 to 12  on sublattice A, and $k_{1/2}=$13 to 24 for sublattice B. 
The 1st outgoing particle $k_3$ is at point 1 and sublattice A. In c), the 2nd outgoing particle is on sublattice A, in d) on sublattice B. }
 \label{ThexUVEx} 
\end{figure}

In Fig.~\ref{ThexUVEx} we display the flow of various susceptibilities for SDW and CDW instabilities, and $V_T$ very close to $T_c$. The actual calculation takes place in the fermionic basis which diagonalizes the hopping term. The resulting interactions are transformed back into the sublattice basis with operators $c^{(\dagger)}_{\vec{k},s,b}$ on sublattice $b=A$,$B$ for incoming and outgoing particles. In Fig.~\ref{ThexUVEx} c), d) we show the real part of the effective $V_T(k_1,k_2,k_3)$ for a small number of $N=12$ $\vec{k}$-space points near the Dirac points (with points 1,2 and $\vec{K}$, 3 and 4 at $\vec{K}'$, continuing clockwise around the Brillouin zone hexagon) close to $T_c$ of the SDW instability. Patch indices 1 to 12 belong to particles on sublattice $A$, and 13 to 24 to sublattice $B$. As function of the incoming $k_1$, $k_2$ with outgoing $k_3$ fixed, $V_T(k_1,k_2,k_3)$ shows either vertical or horizontal features with strongly attractive or repulsive values. The vertical features have $k_2=k_3$ (i.e. same wavevector, same sublattice for particles 2 and 3) or $k_2=k_3\pm 12$ (same wavevector, but different sublattices for particles 2 and 3). 
We can compare $V_T(k_1,k_2,k_3)$ with an infinite-range interaction which gives a SDW as groundstate. On a lattice with $N$ sites, we can define spin-spin interactions 
$ N^{-1} \sum_{\vec{q}} J_{\vec{q}}^{b,b'} \vec{S}_{\vec{q}}^{b} \cdot \vec{S}_{-\vec{q}}^{b'}$ with $\vec{S}_{\vec{q}}^{b} = \frac{1}{2} \sum_{\vec{k}}\vec{\sigma}_{ss'} c^{\dagger}_{\vec{k}+\vec{q},s,b}c_{\vec{k},s',b}$. For the infinite-range SDW interaction, only the $\vec{q}=0$ components in $J_{\vec{q},b,b'}$ are nonzero. We should have $J_{\vec{q}=0}^{b,b}<0$, i.e. ferromagnetic (FM) on the same sublattice, while $J_{\vec{q}=0}^{b,b'}>0$, i.e. AF, for different sublattices. 
Comparing this with the effective interaction from the fRG, \[ \frac{1}{2N} \sum V_T^{b_1b_2b_3} (\vec{k},\vec{k}',\vec{k}+\vec{q}) c^{\dagger}_{\vec{k}+\vec{q},s,b_3}c^{\dagger}_{\vec{k}'-\vec{q},s',b_4} c_{\vec{k}',s',b_2} c_{\vec{k},s,b_1} , \] we get $V_T^{bb'b'} (\vec{k},\vec{k}',\vec{k}'-\vec{q})=-J_{\vec{q}}^{b,b'}$ and  $V_T^{bb'b} (\vec{k},\vec{k}',\vec{k}+\vec{q})=-J_{\vec{q}}^{b,b'}/2$. 
In the fRG data in Fig.~\ref{Thexplot} c) and d), only the $\vec{q}=0$ interactions grow strongly, and the signs depending on the sublattice follow exactly that of the reduced spin-spin-interaction with FM intra-sublattice and AF inter-subband processes.
The CDW instability can be read off from $V_T$ in a similar way.
 
\begin{figure}

\includegraphics[width=.48\textwidth]{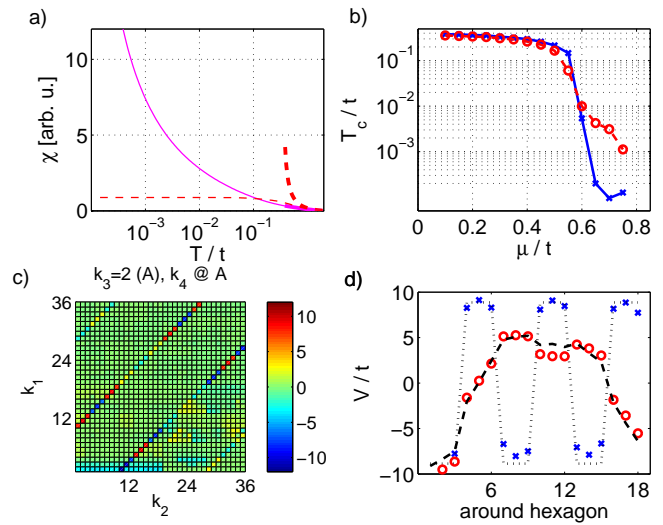}

\caption{(color online). 
a) Flow of pairing susceptibilities in $f$-wave (solid lines) and CDW channel (dashed lines) for chemical potentials $\mu=0.1t$ (thick lines) to $\mu=0.75t$ (thin lines).  
b) Critical temperatures for the flow to strong coupling vs. $\mu$. Crosses: $U$$=$$1.2t$, $V$$=$$2.4t$ with a CDW instability for $\mu<0.7t$ and a triplet Cooper instability for $\mu>0.7t$. Circles: $U$$=$$1.2t$, $J$$=$$2.4t$ with a SDW instability for $\mu<0.65t$ and a singlet $d$-wave Cooper instability for $\mu \ge 0.65t$. 
c) Effective interaction at low scales for $U=1.2t$, $V=2.4t$, $\mu=0.75t$, and outgoing wavevectors fixed on sublattice $A$, $k_3=2$ for a $18\times 3$ discretization. Points $k_1$ and $k_2$ are on the middle of the three rings nearest to the FS and on sublattice $A$ for index 1 to 18 and on B for 19 to 36. The colorbar indicates the values of the couplings. 
d) Crosses: Pair scattering $V_T (k_1,\bar{k}_1, k_2)$  for $U$$=$$1.2t$, $V$$=$$2.4t$, $\mu=0.75t$ with incoming and outgoing particles on sublattice $A$ and $\vec{k}_1+\vec{k}(\bar{k}_1)=0$ vs. $k_2$ around the Brillouin zone 
hexagon. 
The dotted line is the ansatz $-V_f d_f^*(\vec{k}_1)d_f(\vec{k}_2)$.  
Circles: Same data for the $d$-wave instability at $U=1.2t$, $J=2.4t$, $\mu=0.75t$. Here the dashed line is $-V_d [ d_{x^2-y^2}^* (\vec{k}_1) d_{x^2-y^2} (\vec{k}_2)
+   d_{xy}^* (\vec{k}_1) d_{xy} (\vec{k}_2)]$. 
 } 
\label{pairplot}
\end{figure} 

Next we turn to the doped system. Moving $\mu$ away from zero, we obtain two FSs around the Dirac points and the $\vec{q}=0$ nesting between the two bands is reduced. 
This cuts off the CDW and SDW instabilities at low $T$. The behavior of the CDW susceptibility and $T_c$ vs. $\mu$ is shown in Fig.~\ref{pairplot} a), b). The SDW instability for dominant $U$ (or $J$) behaves analogously. 
Beyond a critical doping, the CDW susceptibility remains finite for $T \to 0$. If we continue the flow down to lower temperatures  $T< 10^{-3}t$, we observe a strong growth in the Cooper pairing processes with zero total incoming wavevector. This is clearly visible in the effective interactions near $T_c$ shown in Fig.~\ref{pairplot} c) for $\mu=0.75t$. Here, processes with zero total incoming wavevector (diagonal features) are enhanced strongly. 
The pair scattering $V_T(\vec{k},-\vec{k} \to \vec{k}',-\vec{k}')$ is odd with respect to reversal of the outgoing (or incoming) wavevectors, corresponding to triplet pairing. From Fig.~\ref{pairplot} c) it can inferred that the pair partners $(\vec{k},-\vec{k})$ are on the same sublattice.  The pair scattering between wavevectors near the same Dirac point is attractive, and from one  to the other Dirac point it is repulsive. As shown in Fig \ref{pairplot} d), the zero-total-momentum part of the effective interaction follows closely the form $- V_f d_f(\vec{k})d_f(\vec{k}')$ with the $f$-wave form factor $d_f(\vec{k})= \sin(k_x) - 2\sin (k_x/2) \cos (\sqrt{3} k_y/2)$. The pairing has the same sign on the two sublattices. 
The corresponding meanfield picture gives a nodeless state with gap amplitudes of opposite sign on the two Fermi circles. 
In real space, the pairing with $d_f(\vec{k})$ takes place between a given site and its 6 next-nearest neighbors, with sign change upon a $\pi/3$-rotation around the site. Thus the pairing is not directly mediated by the nearest-neighbor repulsion $V$, but it is due to a next-nearest neighbor attraction that is generated by second and higher orders of $V$ summed up in the fRG flow. 
Consistently,  if we introduce a next-nearest neighbor attraction $V_2<0$, the pairing instability is enhanced. Note that the intra-sublattice pairing found here is quite different from the nearest-neighbor inter-sublattice pairing found in a meanfield study in Ref.  \onlinecite{uchoa} for attractive $V$.
In full agreement with the analysis of the interactions, the flow of the $f$-wave susceptibility with form factor $d_f(\vec{k})$ (see Fig.~\ref{pairplot} a)) shows a strong upturn once the CDW susceptibility is cut off by sufficient doping. Other pairing susceptibilities grow much more weakly.    

The SDW regime driven by the onsite-$U$ does not produce any measurable scale for a superconducting instability within our numerical precision. 
Upon doping, the SDW regime just gives way to a stable Fermi liquid. The SDW regime can also be generated with a antiferromagnetic Heisenberg interaction $J$ on nearest neighbors. The semimetal becomes unstable with respect to the SDW  for $J>2t$. Doping into this state (see Fig.~\ref{pairplot} b)), the SDW scale breaks down. Now we find an instability in the singlet pairing channel with $d$-wave symmetry, that can also be found from a meanfield decoupling of the $J$-term\cite{blackschaffer}. The energetically best state is the complex linear combination of the two degenerate basis function $d_{xy}$ and $d_{x^2-y^2}$, leading to a fully gapped time-reversal-symmetry-breaking state\cite{blackschaffer}. In Fig.~\ref{pairplot} d) we show the effective pair scattering. It follows closely the ansatz $-V_d [ d_{x^2-y^2}^* (\vec{k}) d_{x^2-y^2} (\vec{k}')
+   d_{xy}^* (\vec{k}) d_{xy} (\vec{k}')]$ with the form factors corresponding to next-nearest neighbor pairing of $d_{x^2-y^2}$- and $d_{xy}$-type, $d_{x^2-y^2}(\vec{k}) = e^{-i k_y/\sqrt{3}}-  e^{i k_y/(2\sqrt{3})} \cos k_x/2 $ and $d_{xy}(\vec{k})= i e^{i k_y/(2\sqrt{3})}\sin k_x/2$.
 
Regarding possible realizations of Cooper pairing, we note that a sizable $T_c$ for the triplet pairing requires dominant nearest-neighbor repulsion $V>U$ large enough to be at least close to a CDW ordered state for zero doping. $V>U$ could be realized due to Holstein phonons, reducing the effective $U$\cite{sangiovanni}. A CDW state (or an antiferromagnetic insulator indicative of strong Heisenberg exchange $J$ which could lead to robust $d+id$ pairing), however, does seem to be realized in graphene. 
Yet, for graphene on a substrate, there have been theoretical ideas\cite{lederer} that out-of-plane vibrations of the carbon atoms with opposite amplitude on the two sublattices could trigger a CDW instability, at least in a magnetic field. If the system is near such an instability, the phononic fluctuations would add to the intrinsic fluctuations, and move the system closer to the parameter range for triplet pairing. Recent photoemission work\cite{zhou} for graphene on SiC substrates revealed a gap-like feature near the Dirac points, interpreted as AB-sublattice symmetry-breaking by the substrate. If this gap is in fact a cooperative effect of substrate and electronic interactions, it would seem promising to dope the system out of  gapped phase and to search for superconducting correlations.

In conclusion, we have analyzed instabilities of interacting electrons on the honeycomb lattice using a fRG method. The undoped state becomes unstable with respect to spin-density wave and charge-density wave instabilities, if onsite or nearest-neighbor repulsions exceed critical values. Upon sufficient doping, the CDW instability gives way to a triplet-pairing instability with intra-sublattice Cooper pairing of next-nearest neighbors.  Doping of a SDW regime with $J>0$ leads to a singlet-pairing instability in the $d$-wave channel.

CH thanks A. Black-Schaffer, G.H. Gweon and M. Kinza for discussions and BaCaTec for financial support.

\end{document}